\documentclass[9pt, onecolumn]{article}
\usepackage{times,bm}
\usepackage[affil-it,]{authblk}
\usepackage[pagebackref=false,colorlinks=true,linkcolor=olive,citecolor=red,urlcolor=black]{hyperref}
\usepackage{geometry}
\usepackage[switch*, displaymath,pagewise, mathlines]{lineno}
\usepackage{graphicx}
\usepackage{cite}
\usepackage{amsmath}
\usepackage{amsfonts}
\usepackage{amssymb}
\usepackage{subfloat}
\usepackage{slashed}
\usepackage{color}
\usepackage{multirow,multicol}
\usepackage{tabulary}
\usepackage[flushleft]{threeparttable}
\usepackage{pgfplots,subfigure}
\usepackage{xcolor}
\numberwithin{equation}{section}
\usepackage{tikz}
\usepackage{lipsum}
\usepackage{ulem}
\usepackage{hyperref}
\usepackage{nameref}
\usepackage{comment}

\usepgflibrary{arrows}
\usetikzlibrary{shapes.callouts}
\tikzset{
	level/.style   = { thick, },
	connect/.style = { dotted, red   },
	notice/.style  = { draw, rectangle callout, callout relative pointer={#1} },
	label/.style   = { text width=2cm }
}
\definecolor{acsblue}{RGB}{39,117,215}
\definecolor{shadecolor}{RGB}{255,241,204}
\usepackage{letltxmacro}
\makeatletter
\let\oldr@@t\r@@t
\def\r@@t#1#2{%
	\setbox0=\hbox{$\oldr@@t#1{#2\,}$}\dimen0=\ht0
	\advance\dimen0-0.2\ht0
	\setbox2=\hbox{\vrule height\ht0 depth -\dimen0}%
	{\box0\lower0.4pt\box2}}
\LetLtxMacro{\oldsqrt}{\sqrt}
\renewcommand*{\sqrt}[2][\ ]{\oldsqrt[#1]{#2}}
\makeatother
\usepackage{environ}

\begin{document}

\newcommand{{\ri}}{{\rm{i}}}
\newcommand{{\Psibar}}{{\bar{\Psi}}}

\title{\mdseries{Fermion-antifermion pairs in magnetized spacetime generated by a point source}}
\author{ \textit {\mdseries{Abdullah Guvendi}}$^{\ 1}$\footnote{\textit{ E-mail: abdullah.guvendi@erzurum.edu.tr (Corresp. Auth.)} }~,~ \textit {\mdseries{Omar Mustafa}}$^{\ 2}$\footnote{\textit{ E-mail: omar.mustafa@emu.edu.tr} }   \\
	\small \textit {$^{\ 1}$ Department of Basic Sciences, Erzurum Technical University, 25050, Erzurum,
		Turkiye}\\
	\small \textit {$^{\ 2}$ Department of Physics, Eastern Mediterranean University, G. Magusa, north Cyprus, Mersin 10 - Turkiye }}

\date{}
\maketitle

\begin{abstract}
In this research, we study fermion-antifermion pairs in a magnetized spacetime induced by a point-like source and characterized by an angular deficit parameter, \(\alpha\). In the rest frame, the relative motion (\(\propto r\)) of these pairs is analyzed using exact solutions of a two-body Dirac equation with a position-dependent mass expressed as \(m(r) = m_0 + \mathcal{S}(r)\). We select the Lorentz scalar potential \(\mathcal{S}(r) = -\alpha_c/r\), which modifies the rest mass in a manner analogous to an attractive Coulomb potential, and derive analytical solutions to the resulting radial wave equation. Our findings are applicable to pairs in flat spacetime when \(\alpha = 1\) without loss of generality. We elucidate how the spectra of such pairs are influenced by the spacetime background. Additionally, we observe that even the well-known non-relativistic energy (\(\propto \alpha_c^2\)) reflects the influence of the parameter \(\alpha\) in positronium-like fermion-antifermion systems. We propose that our results can also be extended to study charge carriers in magnetized monolayer materials. Furthermore, we demonstrate that the metric for a 2+1-dimensional spinning point source background can be transformed into the metric describing the near-horizon region of a rotating BTZ black hole, a result not previously reported in the literature. This metric holds potential for providing meaningful insights into topics such as holographic superconductivity and quantum critical phenomena in future research.
\end{abstract}

\begin{small}
\begin{center}
\end{center}
\end{small}



\section{Introduction}\label{sec1}

Modern theoretical physics suggests that vacuum phase transitions in the primordial universe may give rise to topological defects, such as cosmic strings, domain walls, and monopoles \cite{1,2,3,4,5,6}. For instance, cosmic strings are one-dimensional defects in spacetime, where, locally, an idealized cosmic string creates a flat region, despite the overall conical geometry of the manifold. These formations are a consequence of symmetry breaking during phase transitions in theories that extend beyond the Standard Model, especially in grand unified theories \cite{4,5,6,7,8}. Due to their high energy per unit length, cosmic strings have the potential to generate observable phenomena, including gravitational waves (GWs) and gravitational lensing. Recent research using Planck data has placed tight limits on the existence of cosmic strings and other topological defects \cite{9,10}. Investigations into gravitational waves generated by cosmic string networks in the early universe have utilized a range of models, including the Abelian Higgs model and the Nambu–Goto action, extending beyond the conventional Nambu–Goto approximation. The search for stochastic gravitational wave backgrounds from cosmic strings has been active for over 15 years, with data from the Parkes Pulsar Timing Array being employed \cite{11}. Moreover, studies have explored the spectrum of neutrinos emitted by cosmic string loops, particularly phenomena such as quasi-cusps, kinks, and kink–kink collisions, as shown by Creque-Sarbinowski et al. \cite{12}. The study of cosmic string spacetime continues to be a key focus of research \cite{13,14,15,16,17,18,19,20}. For example, some studies have investigated Abelian strings in static de Sitter backgrounds \cite{21}, examined the electric and magnetic self-forces due to the conical structure of cosmic string spacetime \cite{22}, and explored the behavior of quantum particles in the presence of topological defects within Kaluza–Klein theory \cite{23}. Other important contributions include the study of interactions between the magnetic quadrupole moments of neutral particle systems and radial electric fields in rotating frames for non-relativistic particles \cite{24}, the derivation of exact black hole spacetime metrics in Einstein–Hilbert-Bumblebee gravity around global monopole fields \cite{25} (see also \cite{S4,S5}), the investigation of evolving topologically deformed wormholes supported by dark matter halos \cite{26}, and the analysis of the influence of cosmological constants and the internal energy densities of cosmic strings on the deflection of light in rotating cosmic string spacetimes with internal structure \cite{27}. Furthermore, studies have addressed the effects of Lorentz symmetry breaking on light deflection in rotating cosmic string spacetimes under weak field approximations (see also \cite{S2}) \cite{28}, as well as the bending of light in regular black holes spacetime with cosmic strings in the weak field limit \cite{29}.

In the relativistic context, the dynamics of two particles' interaction are governed by equations that include both free Hamiltonians and phenomenologically defined interaction potentials. A central challenge in this domain lies in selecting suitable interaction potentials. Additionally, determining the relativistic dynamics of two interacting particles introduces the so-called two-time problem. Heisenberg and Pauli were the first to propose solutions to this issue, paving the way for the development of the initial relativistic two-body equation by Breit \cite{30}. Breit’s approach involved treating electron velocities as Dirac matrices and incorporating them into the Lienard-Wiechert potentials, which allowed him to derive first-order corrections to the Coulomb potential within the Darwin Lagrangian. This work underscored the fundamental connection between electron motion and spin in a relativistic context. However, Breit's equation becomes less efficient at high particle velocities or large distances due to retardation effects. Another approach, introduced by Bethe and Salpeter, proved effective for bound-state problems such as hydrogen-like atoms but led to the many-time problem \cite{31}. The relativistic two-body equation proposed by Kemmer, Fermi, and Yang, though phenomenological, holds significant theoretical importance. Similarly, a fully covariant one-time many-body Dirac equation, derived from quantum electrodynamics via the action principle, encompasses the most complete electric and magnetic potentials \cite{32}. In four-dimensional space, this equation becomes a $16\times16$ matrix equation, with group theoretical methods assisting in the covariant separation of relative and center-of-mass coordinates. Despite its complexity, the hydrogen-like atom energy spectrum can be derived perturbatively from the coupled equations \cite{33}. However, it has been shown that the many-body Dirac equation can yield exact solutions in both flat and curved spaces, particularly for coupled pairs exhibiting dynamical symmetries \cite{34,35,36,37,38,39,40}. Here, we will investigate the influence of an external magnetic field \cite{40} and the spacetime background generated by a static point source (as discussed in \cite{41}) on the relative motion of fermion-antifermion pairs within the framework of position-dependent mass \cite{42,AO-}. In 2 + 1 dimensions, solutions to Einstein's equations for both static single-body and many-body sources can be derived using geometric and analytical methods \cite{41,43,44,45}. For a single body, it has been demonstrated that the surface of the body is situated outside the origin \cite{43,44,45}. The spacetime backgrounds generated by stationary spinning sources and static point sources were initially introduced in 2 + 1 dimensions through these solutions \cite{41,43,44,45} and later extended to 3 + 1 dimensions while preserving cylindrical symmetry \cite{43,44,45}. In this context, the dynamics of a vector field remain invariant under Lorentz boosts along the additional third spatial coordinate (the z-axis in cylindrical coordinates). In 2 + 1 dimensions, the source is located at the spatial origin, where the string intersects the space-like subspace. The 2 + 1 dimensional cosmic string-induced geometric backgrounds, also known as point source-induced spacetime backgrounds \cite{43,44,45}, introduce non-trivial topologies (see also \cite{S1,S3}). While an induced spacetime may appear locally flat, its global characteristics differ, allowing entities such as cosmic strings or point sources to influence physical systems by altering spacetime symmetry. The study of \(f\overline{f}\) pairs can incorporate various interaction potentials, particularly within the framework of the Dirac equation. To account for external fields, two primary approaches are utilized: electromagnetic minimal coupling and the Lorentz scalar potential. Electromagnetic minimal coupling establishes an interaction between the Dirac field and the electromagnetic field via the vector potential \(\mathcal{A}_\mu = (\phi, \mathbf{A})\), where \(\phi\) is the scalar potential and \(\mathbf{A}\) represents the vector potential. In this approach, the partial derivative \(\partial_\mu\) in flat spacetime is replaced by the covariant derivative \(D_\mu = \partial_\mu + ie\mathcal{A}_\mu\) \cite{36,37,39,40}, with \(e\) denoting the electric charge. This substitution preserves gauge invariance and describes the interactions between charged particles and the electromagnetic field, as formulated in quantum electrodynamics. On the other hand, the Lorentz scalar potential alters the Dirac equation by modifying the mass term directly. In this case, the mass \(m\) is replaced with \(m + \mathcal{S}(\mathbf{x})\), where \(\mathcal{S}(\mathbf{x})\) represents the scalar potential \cite{41}. Unlike minimal coupling, this method does not affect the gauge invariance of the Dirac field but instead modifies the particle’s effective mass locally \cite{14}. The primary distinction between these approaches lies in their effects: electromagnetic minimal coupling adjusts the kinetic term through the inclusion of the vector potential in the derivative, thereby modeling interactions with the electromagnetic field. In contrast, the Lorentz scalar potential directly alters the mass term, changing the particle's intrinsic energy \cite{14,41}.

This study investigates the relative motion of coupled fermion-antifermion pairs in a magnetized spacetime induced by a static point source, with a particular emphasis on cases involving position-dependent mass. Assuming that the center of mass remains fixed at the spatial origin, we derive analytical solutions to the covariant many-body Dirac equation governing the system. The structure of the paper is as follows. Section \ref{sec2} outlines the derivation of the general wave equation for the system, while Section \ref{sec3} focuses on obtaining exact solutions under various conditions. Finally, Section \ref{sec4} concludes with a comprehensive discussion of the findings and their implications.

\section{\label{sec2}{Wave equation}} 

In this section, we derive the covariant many-body Dirac equation for a general \( f\overline{f} \) pair, taking into account the effects of electromagnetic and Lorentz scalar potentials within a (2+1)-dimensional cosmic string spacetime, referred to as the static point source spacetime \cite{41,43,44,45}. The spacetime is characterized by a metric with a negative signature \((+, -, -)\): \[ ds^2 = c^2 dt^2 - dr^2 - \alpha^2 r^2 d\phi^2, \] where \(c\) represents the speed of light and \(0 < \alpha  \leq 1\) is a parameter related to the string tension, proportional to \(G\mu_{s}/c^2 \geq 0\), with \(\alpha\) defined as \(1 - 4G\mu_{s}/c^2\). In the context of general relativity, such a topological defect is characterized by an angular deficit parameter \(\alpha\). For \(0 < \alpha < 1\), the resulting spacetime takes on a conical topology, with its apex located at \(r = 0\) \cite{41,43,44,45}. This type of spacetime structure can also be used as a model to study the dynamics of charge carriers in a gapped graphene layer \cite{15}. The spacetime of a monolayer graphene sheet can be described within a \(2+1\)-dimensional framework that features constant negative Gaussian curvature \cite{15}. This is attributed to the presence of carbon heptagons in the graphene structure, which induce regions of negative curvature. These regions can be effectively modeled by incorporating an additional angular sector (\(\alpha > 1\)) into the monolayer sheet. Within the position dependent mass framework \cite{42,AO-}, the two-body Dirac equation for fermion-antifermion pairs is expressed as follows \cite{35,AO-}:
\begin{equation}
\begin{split}
&\left\lbrace \left[\slashed{\nabla}_{\mu}^{f} +i\tilde{m}(x_{\mu}^{f}) \mathcal{I}_{2}\right] \otimes \gamma^{t^{\overline{f}}}+ \gamma^{t^{f}}\otimes \left[\slashed{\nabla}_{\mu}^{(\overline{f})} +i\tilde{m}(x_{\mu}^{(\overline{f})}) \mathcal{I}_{2}\right] \right\rbrace \Psi(x_{\mu}^{f},x_{\mu}^{\overline{f}})=0,\\
&\slashed{\nabla}_{\mu}^{f (\overline{f})}=\gamma^{\mu^{f (\overline{f})}}\left[\partial_{\mu}^{f (\overline{f})}+i\frac{e_{f(\overline{f})}\mathcal{A}^{f(\overline{f})}_{\mu}}{2\hbar c}-\Gamma_{\mu}^{f (\overline{f})}\right].\label{eq1}
\end{split}
\end{equation}
In this context, \( \tilde{m} \) is defined as \( mc/\hbar \), where \( m \) represents the rest mass of the fermions, \( e \) denotes the electric charge, and \( \mathcal{A}_{\mu} \) refers to the electromagnetic 3-vector potential. The constant \( \hbar \) is the reduced Planck constant. Greek indices are used to denote spacetime coordinates (\( x^{\mu} = t, r, \phi \)), and the bi-local spinor \( \Psi(x_{\mu}^{f}, x_{\mu}^{\overline{f}}) \) is a function of the spacetime position vectors \( x_{\mu}^{f} \) and \( x_{\mu}^{\overline{f}} \) corresponding to the fermion and antifermion, respectively. The identity matrix \( \mathcal{I}_2 \) is a 2x2 matrix, with fermions and antifermions represented by \( f \) and \( \overline{f} \), respectively. In equation (\ref{eq1}), the space-dependent Dirac matrices \( \gamma^{\mu} \) are derived from the relation \( \gamma^{\mu} = e^{\mu}_{(a)} \gamma^{(a)} \), where \( e^{\mu}_{(a)} \) are the inverse tetrad fields and \( \gamma^{(a)} \) are the constant Dirac matrices. These \( \gamma^{(a)} \) matrices can be expressed in terms of the Pauli spin matrices \( (\sigma_{x}, \sigma_{y}, \sigma_{z}) \) \cite{35}
\begin{equation}
\gamma^{(0)} = \sigma_{z}, \quad \gamma^{(1)} = i\sigma_{x}, \quad \gamma^{(2)} = i\sigma_{y},\label{SDM}
\end{equation}
where \(i = \sqrt{-1}\). In a spacetime with a (2+1)-dimensional negative signature, where the Minkowski metric \(\eta_{(a)(b)}\) is expressed as \(\eta_{(a)(b)} = \textrm{diag}(+, -, -)\), the inverse tetrad fields are given by the equation \(e^{\mu}_{(a)} = g^{\mu\tau} e_{\tau}^{(b)} \eta_{(a)(b)}\). Here, \(g^{\mu\tau}\) denotes the contravariant metric tensor, and \(e_{\tau}^{(b)}\) refers to the tetrad fields, which are derived from the relation \(g_{\mu\tau} = e_{\mu}^{(a)} e_{\tau}^{(b)} \eta_{(a)(b)}\), where \(g_{\mu\tau}\) is the covariant metric tensor. The spinorial affine connections \(\Gamma_{\mu}\) in Eq. (\ref{eq1}) are computed using the formula \(\Gamma_{\lambda} = \frac{1}{8} \left[ e^{(a)}_{\nu, \lambda} e^{\tau (a)} - \Gamma_{\nu \lambda}^{\tau} \right] \mathcal{S}^{\mu\nu}\), where \(\mathcal{S}^{\mu\nu} = \frac{1}{2} \left[ \gamma^{\mu}, \gamma^{\nu} \right]\) is the spin operator \cite{35}. The Christoffel symbols \(\Gamma_{\nu \lambda}^{\tau}\) are given by \(\Gamma_{\nu \lambda}^{\tau} = \frac{1}{2} g^{\tau \epsilon} \left[ \partial_{\nu} g_{\lambda \epsilon} + \partial_{\lambda} g_{\nu \epsilon} - \partial_{\epsilon} g_{\nu \lambda} \right]\). From these expressions, we can derive the generalized Dirac matrices (\(\gamma^{\mu}\)) and the non-zero components of the spinorial affine connections (\(\Gamma_{\mu}\)) for Dirac fields as outlined in \cite{13,15}
\begin{equation}
\begin{split}
\gamma^{t^{f(\overline{f})}}=\frac{1}{c}\gamma^{(0)},\quad \gamma^{r^{f(\overline{f})}}=\gamma^{(1)},\quad \gamma^{\phi^{f(\overline{f})}}=\frac{\gamma^{(2)}}{\alpha \, r^{f(\overline{f})}},\quad \Gamma^{f(\overline{f})}_{\phi}=\frac{i\,\alpha}{2}\sigma_{z}.\label{GDM-SC}
\end{split}
\end{equation}
In the affine spin connections \(\Gamma^{f(\overline{f})}_{\phi}\), only one component is non-zero. Hence, we find
\begin{equation}
\gamma^{\phi^{f(\overline{f})}}\Gamma^{f(\overline{f})}_{\phi} = -\frac{i}{2r^{f(\overline{f})}}\sigma_{x}.
\end{equation}
The influence of a uniform external magnetic field, oriented perpendicular to the \((r,\phi)\)-plane, can be incorporated through the angular component of the electromagnetic 3-vector potential. Specifically, the components are given by \(\mathcal{A}^{f}_{\phi} = \frac{\mathcal{B}_{0}}{2\hbar c} r^{2}_{f}\) and \(\mathcal{A}^{\overline{f}}_{\phi} = \frac{\mathcal{B}_{0}}{2\hbar c} r^{2}_{\overline{f}}\), where \(\mathcal{B}_{0}\) represents the magnitude of the applied magnetic field \cite{40}. To provide a more detailed description, we explicitly present the associated two-body Dirac equation as \(\hat{M}\Psi = 0\), with \(\hat{M}\) denoting
\begin{equation}
\begin{split}
&\gamma^{t^{f}}\otimes\gamma^{t^{\overline{f}}}\left[\partial_{t}^{f}+\partial_{t}^{\overline{f}} \right]+\gamma^{r^{f}}\partial_{r}^{f}\otimes \gamma^{t^{\overline{f}}}+ \gamma^{t^{f}}\otimes \gamma^{r^{\overline{f}}}\partial_{r}^{\overline{f}}+\gamma^{\phi^{f}} \otimes\gamma^{t^{\overline{f}}}\slashed{\partial}_{\phi}^{f}+\gamma^{t^{f}}\otimes \gamma^{\phi^{\overline{f}}}\slashed{\partial}_{\phi}^{\overline{f}}\\
&+i\tilde{m}(x_{\mu}^{f},x_{\mu}^{\overline{f}})\left[\mathcal{I}_{2}\otimes \gamma^{t^{\overline{f}}}+\gamma^{t^{f}}\otimes \mathcal{I}_{2}\right]-\left[\gamma^{\phi^{f}}\Gamma_{\phi}^{f}\otimes \gamma^{t^{\overline{f}}}+\gamma^{t^{f}}\otimes \gamma^{\phi^{\overline{f}}}\Gamma_{\phi}^{\overline{f}} \right],\label{eq4}
\end{split}
\end{equation}
where 
\begin{equation*}
\begin{split}
\slashed{\partial}_{\phi}^{f}\rightarrow \partial_{\phi}^{f}+i\frac{e\mathcal{B}_{0}}{2\hbar c}r^{2}_{f},\quad \slashed{\partial}_{\phi}^{\overline{f}}\rightarrow \partial_{\phi}^{\overline{f}}-i\frac{e\mathcal{B}_{0}}{2\hbar c}r^{2}_{\overline{f}},
\end{split}
\end{equation*}
since \(e_{\overline{f}}=-e_{f}\) \cite{37}. Through an in-depth examination of the spacetime interval in question, we reformulate the spacetime-dependent bi-spinor \(\Psi(t, r, R)\) in a separable form. This allows us to express \(\Psi\) as \(e^{-i\omega t} e^{i\vec{K} \cdot \vec{R}} \tilde{\Psi}(\vec{r})\), where  
\[
\tilde{\Psi}(\vec{r}) = e^{i s \phi} (\psi_{1}(r), \psi_{2}(r), \psi_{3}(r), \psi_{4}(r))^T.
\]  
Here, \(\omega\) represents the relativistic frequency, while \(\vec{r}\) and \(\vec{R}\) denote the position vectors for the relative motion and the center of mass, respectively. The vector \(\vec{K}\) is associated with the momentum of the center of mass, and \(s\) refers to the total spin of the coupled \(f\overline{f}\) system. The superscript \(^T\) indicates the transpose of the spinor that depends on \(r\). Following the conventional framework for two-body systems, we define the relative and center-of-mass coordinates as outlined in \cite{36,37}
\begin{equation}
\begin{split}
&R_{x^{\mu}}=\frac{x^{\mu^{f}}}{2}+\frac{x^{\mu^{\overline{f}}}}{2},\quad r_{x^{\mu}}=x^{\mu^{f}}-x^{\mu^{\overline{f}}}.\quad x^{\mu^{f}}=\frac{1}{2}r_{x^{\mu}}+R_{x^{\mu}},\quad x^{\mu^{\overline{f}}}=-\frac{1}{2}r_{x^{\mu}}+R_{x^{\mu}},\\
&\partial_{x_{\mu}}^{f}=\partial_{r_{x^{\mu}}}+\frac{1}{2}\partial_{R_{x^{\mu}}},\quad \partial_{x_{\mu}}^{\overline{f}}=-\partial_{r_{x^{\mu}}}+\frac{1}{2}\partial_{R_{x^{\mu}}}, \label{eq4-}
\end{split}
\end{equation}
for a $f\overline{f}$ pair. It is crucial to note that the expression \(\partial_{x^{\mu}}^{f} + \partial_{x^{\mu}}^{\overline{f}}\) simplifies to \(\partial_{R_{x^\mu}}\). This reduction indicates that the dynamics of the system, governed by the relativistic frequency \(\omega\), are related to the proper time, denoted by \(\partial_{R_{t}}\). In order to isolate the effects of pairing, we assume the center of mass remains stationary at the spatial origin \cite{15}. This assumption facilitates the development of equations describing the relative motion of the pair in the center of mass reference frame, where the total momentum (\(\hbar \vec{K}\)) is zero \cite{new}. By combining these equations, we obtain an equation set, under the assumption that the total mass of the system depends solely on the relative radial distance \(r\) between the particles \cite{AO-}
\begin{equation}
\begin{split}
&\varpi \xi_{1}(r)-2\tilde{m}(r)\xi_{2}(r)+2\left(\frac{1}{r}+\partial_{r}\right)\xi_{4}(r)-\frac{4\left(s+\mathcal{B}r^2\right)}{\alpha\,r}\xi_{3}(r)=0,\\
&\varpi \xi_{2}(r)-2\tilde{m}(r)\xi_{1}(r)=0,\quad \varpi \xi_{3}(r)-\frac{4\left(s+\mathcal{B}r^2\right)}{\alpha\,r}\xi_{1}(r)=0,\quad \varpi \xi_{4}(r)-2\left(\frac{1}{r}+\partial_{r}\right) \xi_{1}(r)=0,\label{eqset}
\end{split}
\end{equation}
where $\varpi=\omega/c$, $\mathcal{B}=\frac{e\mathcal{B}_{0}}{8\hbar c}$, and
\begin{equation}
\begin{split}
&\xi_{1}(r)=\psi_{1}(r)+\psi_{4}(r),\quad \xi_{2}(r)=\psi_{1}(r)-\psi_{4}(r),\\
&\xi_{3}(r)=\psi_{2}(r)+\psi_{3}(r),\quad \xi_{4}(r)=\psi_{2}(r)-\psi_{3}(r).
\end{split}
\end{equation} 
These equations lead to the following wave equation
\begin{equation}
\partial^{2}_{r}\xi_{1}+\frac{2}{r}\partial_{r}\xi_{1}+\left[\frac{\varpi^2-m^{2}_{\dagger}(r)}{4}-\frac{4\left(s+\mathcal{B}r^2\right)^{2}}{\alpha^2\,r^2}\right]\xi_{1}=0,\label{WE}
\end{equation}
where $m_{\dagger}(r)=2\tilde{m}+\mathcal{S}(r)$. It is crucial to note that this equation can be examined by making specific mass modifications tailored to the systems under consideration, as will be explained in the upcoming sections. Therefore, we contend that our model offers a solid foundation for a range of future studies. We are now ready to demonstrate how our model functions.

\section{Exact solutions}\label{sec3}

In this section, we explore exact solutions to the wave equation for various scenarios. Specifically, in section \ref{sec3.1}, we examine the case where \(S(r) = -\alpha_{c}/r\) and \(\mathcal{B}_{0} = 0\). In section \ref{sec3.2}, we investigate the case where \(S(r) = -\alpha_{c}/r\), \(\mathcal{B}_{0} \neq 0\), and \(m = 0\).

\subsection{Exact Solutions for \( S(r) = -\alpha_{c}/r \) and \(\mathcal{B}_{0} = 0\)}\label{sec3.1}

\begin{figure}
\centering
\includegraphics[scale=0.60]{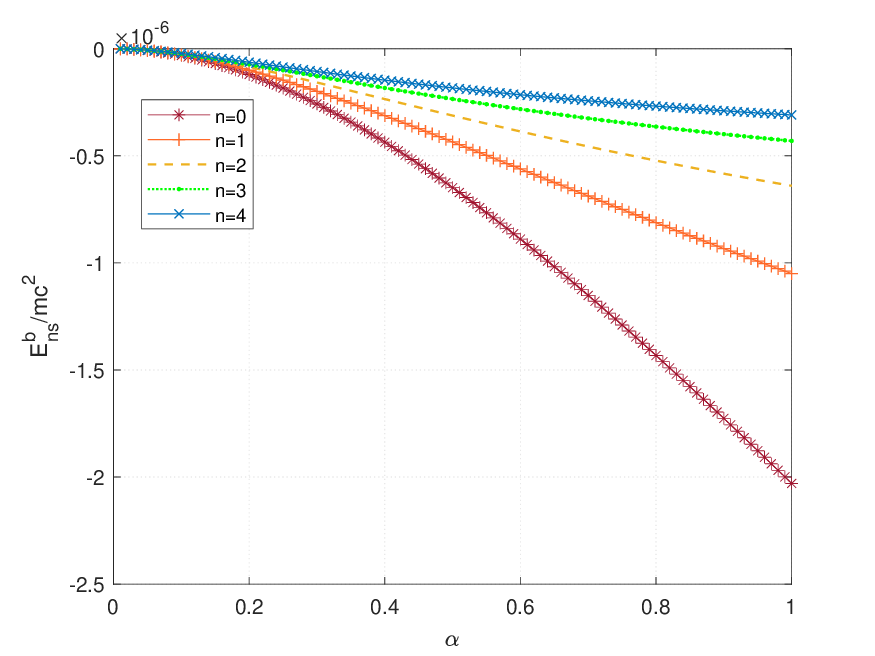}
\caption{Binding energy $\mathcal{E}^{b}_{ns}(\alpha)$ for varying $\alpha$ values. Here, we take $\alpha_{c}=1/137$ and $s=1$.}
\label{fig:1}
\end{figure}

When \(\mathcal{S}(r) = -\alpha_{c}/r\), with \(\alpha_{c}\) being the fine structure constant, Equation (\ref{WE}) can be re-expressed by employing a new dimensionless variable \(z = \sqrt{(2\tilde{m})^2 - \varpi^2}\, r\). This change of variables, along with the ansatz \(\xi_{1}(z) = \frac{1}{z}\xi(z)\), would yield
\begin{equation}
\xi^{''}(z)+\left[\frac{\tilde{a}}{z}-\frac{1}{4}+\frac{\frac{1}{4}-\tilde{b}^2}{z^2}\right]\xi(z)=0,\label{Whit}
\end{equation}
where
\begin{equation*}
\tilde{a}=\frac{2\tilde{m}\alpha_{c}}{2\sqrt{(2\tilde{m})^2-\varpi^2}},\quad \tilde{b}=\frac{\sqrt{1+\alpha_{c}^2+16\tilde{s}^2}}{2},\quad \tilde{s}=\frac{s}{\alpha}.
\end{equation*}
The solution to Equation (\ref{Whit}) can be expressed using the Confluent Hypergeometric function as \(\xi(z) = e^{-\frac{z}{2}}z^{\frac{1}{2}+\tilde{b}} \ _1F_{1}(\frac{1}{2}+\tilde{b}-\tilde{a}, 1+2\tilde{b}; z)\) near the regular singular point at \(z = 0\). It is important to note that this analysis pertains to a bound pair system. To ensure that \(\xi(z)\) remains finite and square integrable, we must truncate the Confluent Hypergeometric series to a polynomial of order \(n \geq 0\). This is done by taking \(\frac{1}{2} + \tilde{b} - \tilde{a} = -n\) \cite{35}. As a result, we obtain the quantization condition (\(\omega \rightarrow \omega_{ns}\)) for such a bound pair, which results in
\begin{equation*}
  \hbar\omega_{ns}=\pm 2m c^2\sqrt{1-\frac{\alpha_{c}^2}{(2n+1+\sqrt{1+\alpha_{c}^2+16\tilde{s}^2})^2}}.  
\end{equation*}
Thus, the energies can be expressed through the power series
\begin{equation}
\begin{split}
&\qquad \mathcal{E}_{ns}\approx 2mc^2 \left\lbrace 1-\frac{\alpha_{c}^2}{2\,(2n+1+\sqrt{1+16\tilde{s}^2})^2}+\frac{(8n+4+3\sqrt{1+16\tilde{s}^2})\,\alpha_{c}^4}{8\,(2n+1+\sqrt{1+16\tilde{s}^2})^4\,\sqrt{1+16\tilde{s}^2}}-\mathcal{O}(\alpha_{c}^{6})\right\rbrace\\
&\Rightarrow \mathcal{E}_{n0}\approx 2mc^2 \left\lbrace 1-\frac{1}{8}\frac{\alpha_{c}^2}{(n+1)^2}+\frac{1}{128}\frac{\alpha_{c}^4}{(n+1)^4}-\mathcal{O}(\alpha_{c}^6)\right\rbrace.\label{eq7-}
\end{split}
\end{equation}
The latest expression provides the familiar non-relativistic binding energy \((\mathcal{E}^{b}_{n0})\) as follows:
\begin{equation*}
\mathcal{E}^{b}_{n0} \approx -\frac{mc^2}{4} \frac{\alpha_{c}^2}{(n+1)^2}.
\end{equation*}
This result pertains to singlet positronium in its ground state (\(n=0\)), where \(\alpha_{c}\) denotes the fine structure constant (approximately \(1/137\)), and \(m\) represents the mass of the electron or positron. For such a system, \(\mathcal{E}^{b}_{00} \approx -mc^2 \frac{\alpha_{c}^2}{4} = -6.803 \text{ eV}\). In cases where \(s \neq 0\), the energy spectrum (\ref{eq7-}) reveals that even the non-relativistic energy, which scales with \(\alpha_{c}^2\), is influenced by spacetime topology, as represented by \(\tilde{s} = s / \alpha\). However, for singlet positronium-like systems, the effect of spacetime topology is not apparent because it is embedded in the total spin of the composite system formed by the coupled fermion-antifermion pair. The effect of spacetime topology on binding energy levels is depicted in Figure \ref{fig:1}. Moreover, it is clear from the energy spectra described by Eq. (\ref{eq7-}) that they consist exclusively of \(s^2\) terms, indicating that the \(s = \pm 1\) quantum states are degenerate when \(\mathcal{B}_{0} = 0\).

\subsection{Exact results in case where $S(r)=-\alpha_{c}/r$, $\mathcal{B}_{0}\neq0$ and $m=0$}\label{sec3.2}

When $S(r)=-\alpha_{c}/r$, $\mathcal{B}_{0}\neq0$ and $m=0$, the wave equation simplifies to:
\begin{equation}
\partial^{2}_{r}\xi_{1} + \frac{2}{r} \partial_{r}\xi_{1} + \left[ \frac{\varpi^2}{4} - \frac{\alpha_{c}^{2}}{4r^2} - \frac{4 \left( s + \mathcal{B}r^2 \right)^{2}}{\alpha^2 r^2} \right] \xi_{1} = 0. \label{WE-new}
\end{equation}
By introducing a new variable \( y = r^2 \frac{2\mathcal{B}}{\alpha} \) and applying the ansatz \( \xi_{1}(y) = y^{-3/4} \xi(y) \), the wave equation becomes:
\begin{equation}
\xi''(y) + \left[ \frac{\tilde{c}}{y} - \frac{1}{4} + \frac{\frac{1}{4} - \tilde{d}^2}{y^2} \right] \xi(y) = 0, \label{Whit-1}
\end{equation}
where
\begin{equation*}
\tilde{c} = \frac{\alpha^2 \varpi^2 - 32 \mathcal{B} s}{32 \alpha \mathcal{B}}, \quad \tilde{d} = \frac{\sqrt{4 + \alpha_{c}^2 + 16 \tilde{s}^2}}{4}.
\end{equation*}
Following the approach detailed in section \ref{sec3.1}, the quantization condition for the energy of this system is given by \( \tilde{c} = n + \frac{1}{2} + \tilde{d} \) where $n=0,1,2...$ \cite{35}. This leads to the energy spectrum:
\begin{equation}
\mathcal{E}_{ns}(\alpha) = \pm 2 \frac{\hbar c}{\ell_{\mathcal{B}}} \sqrt{\alpha} \left[ n + \frac{1}{2} +\tilde{s} + \frac{\sqrt{4 + \alpha_{c}^2 + 16 \tilde{s}^2}}{4} \right]^{1/2}. \label{spec-grap}
\end{equation}
Here, \( \ell_{\mathcal{B}} = \sqrt{\frac{\hbar c}{e \mathcal{B}_{0}}} \) represents the magnetic length \cite{new-1}. This formula applies to charge carriers interacting within magnetized monolayer materials, where \( c \) is replaced by \( v_{F} \) and \( \alpha_{c} \) is replaced by \( \alpha_{c} / \epsilon_{eff} \). In this context, \( v_{F} \) is the Fermi velocity (approximately \( c/300 \)), and \( \epsilon_{eff} \) is the effective dielectric constant of the surrounding medium \cite{new-2,NEW}. When \( \alpha_{c} = 0 \), \( \alpha > 1 \), and \( s = 0 \), the energy spectrum simplifies to:
\begin{equation}
\mathcal{E}_{n}(\alpha) = \pm 2 \frac{\hbar v_{F}}{\ell_{\mathcal{B}}} \sqrt{\alpha} \sqrt{n + 1}. \label{spec-grap-2}
\end{equation}
The energy spectrum in question corresponds to the Landau levels of a spinless Weyl pair (non-interacting) within a monolayer graphene sheet (see also \cite{46,47}). It is important to highlight that the spacetime of monolayer graphene can be represented as a (2 + 1)-dimensional space characterized by constant negative curvature. This arises because the atomic structure of graphene, often illustrated as a carbon hexagon, can be modeled by an additional angular sector (\(\alpha>1\)) due to its intrinsic negative curvature. On the other hand, the spatial components of line elements that describe cosmological topological defects, which have constant positive curvature, are analogous to those encountered in solid-state systems. These topological defects exhibit an angular deficit that is proportional to \(\alpha\), where \(0<\alpha\leq1\), influencing the underlying geometry. This framework enables the investigation of Weyl particle dynamics in an anti-conical spacetime, considering scenarios where \(\alpha > 1\). Furthermore, the results are applicable to the flat monolayer background when \(\alpha = 1\), ensuring the generality and broad relevance of the findings. \\

On the other hand, the metric describing a \((2+1)\)-dimensional spacetime background generated by a spinning point source \cite{41,48}, is given by
\begin{equation}
ds^2=(dt+\tilde{\varpi}d\phi)^2-d\rho^2-(\eta^2\rho^2-\tilde{\varpi}^2)d\phi^2,\label{RSS}
\end{equation}
where \(c=1=\hbar=G\), features a negative signature. Here, \(\tilde{\varpi}\) is related to the angular momentum density of the rotating string source \cite{41,48}, and \(\eta\) (with \(0<\eta<1\)) represents the string tension or angular deficit in the background. By making the identification \(\phi\rightarrow -\frac{i}{\tilde{\alpha}\ell}\,t\), the metric (\ref{RSS}) transforms into
\begin{equation}
ds^2=\left(\frac{\tilde{\alpha}^2\rho^2}{\ell^2}-J^2\right)dt^2-d\rho^2-(\alpha\ell\, d\phi-J\, dt)^2,\label{NHRB}
\end{equation}
where \(\tilde{\alpha}^2=\eta\) (so \(0<\tilde{\alpha}^2<1\)) is associated with the black hole mass, \(\ell\) is related to the cosmological constant, and \(J=\frac{\tilde{\varpi}}{\alpha\ell}\) represents the angular momentum of the rotating BTZ black hole. Notably, when \(J=0\), the metric (\ref{NHRB}) simplifies to
\begin{equation}
ds^2=\frac{\tilde{\alpha}^2\rho^2}{\ell^2}dt^2-d\rho^2-\alpha^2\ell^2 d\phi^2,\label{NHB}
\end{equation}
which, as detailed previously, describes the near-horizon region of the static BTZ black hole \cite{35,49,50,51,52}. This solution (\ref{NHB}), termed the "particle" solution, exhibits conical singularities at \(\rho = 0\) without a horizon, with a deficit angle \(\Omega = 2\pi(1 - \tilde{\alpha})\) \cite{35,49}. When \(\tilde{\alpha} = 1\), the metric corresponds to a singularity-free global \(AdS_{3}\) spacetime \cite{49}. Therefore, the transformation \(\phi\rightarrow -\frac{i}{\tilde{\alpha}\ell}\,t\) shifts the defect from the one surface to another surface. Holographic superconductivity utilizes the \(AdS_{d}/CFT_{d-1}\) correspondence to model superconducting phases by translating them into higher-dimensional gravitational theories, allowing for detailed investigations of strongly correlated electron systems and quantum phase transitions. Quantum critical points, marked by zero-temperature or zero-energy phase transitions driven by quantum fluctuations rather than thermal effects, reveal intricate phenomena such as non-Fermi liquid states. The integration of these concepts facilitates a deep understanding of how superconductivity and quantum critical phenomena interact, providing valuable insights into the behavior of materials near critical points and in strongly interacting regimes. Thus, we propose that the metrics given by Eqs. (\ref{NHRB}) and (\ref{NHB}) could be instrumental in deriving exact results for charged scalar fields and Dirac or Weyl fermions, potentially offering significant contributions to the study of holographic superconductivity and quantum criticality.

\section{Summary and discussions}\label{sec4}

In this study, we investigated the dynamics of fermion-antifermion pairs in a uniform external magnetic field ($\mathcal{B}_{0}$) within a (2+1)-dimensional spacetime, influenced by a point source with an angular deficit parameter \(0<\alpha\leq1\). The primary focus was on the relative motion of these pairs, which we analyzed by solving the many-body Dirac equation with a position-dependent mass (\( m \rightarrow m + \mathcal{S}(r) \)). We adopted the Lorentz scalar potential \(\mathcal{S}(r) = -\alpha_{c}/r\), which modifies the rest mass in a manner analogous to an attractive Coulomb potential. This choice facilitated the derivation of exact analytical solutions for the radial wave equation, without relying on approximations. Our results are valid for fermion-antifermion pairs in both globally and locally flat spacetimes when \(\alpha = 1\), and we computed the non-relativistic ground state binding energy of a singlet ($s=0$) positronium, which was found to be approximately $-6.803$ eV, assuming \(\alpha_{c}=1/137\). We also provided insights into the influence of the magnetized spacetime on the energy spectra of fermion-antifermion pairs under various physical conditions. Notably, for \(s \neq 0\), we observed that even the typical non-relativistic energy (\(\propto \alpha_{c}^2\)) exhibits a dependence on the parameter \(\alpha\) in positronium-like systems when $\mathcal{B}_{0}=0$. Additionally, our findings extend to the investigation of charge carriers in magnetized monolayer materials.

To summarize, we present the exact binding energy levels ($\mathcal{E}^{b}_{ns}$) for positronium-like systems within the point-source induced spacetime background:
\begin{equation}
\begin{split}
&\qquad \mathcal{E}_{ns}\approx 2mc^2 \left\lbrace -\frac{\alpha_{c}^2}{2\,(2n+1+\sqrt{1+16\tilde{s}^2})^2}+\frac{(8n+4+3\sqrt{1+16\tilde{s}^2})\,\alpha_{c}^4}{8\,(2n+1+\sqrt{1+16\tilde{s}^2})^4\,\sqrt{1+16\tilde{s}^2}}\right\rbrace\\
&\Rightarrow \mathcal{E}^{b}_{n0}\approx 2mc^2 \left\lbrace -\frac{1}{8}\frac{\alpha_{c}^2}{(n+1)^2}+\frac{1}{128}\frac{\alpha_{c}^4}{(n+1)^4}\right\rbrace,\label{eq7}
\end{split}
\end{equation}
when $\mathcal{B}_{0}=0$. This expression provides the well-known non-relativistic binding energy ($\propto \alpha_{c}^{2}$) as:
\begin{equation*}
\mathcal{E}^{b}_{n0} \approx -\frac{mc^2}{4} \frac{\alpha_{c}^2}{(n+1)^2},
\end{equation*}
for a spinless pair. This result is applicable to singlet ($s=0$) positronium in its ground state ($n=0$), with \(\alpha_{c}\) as the fine structure constant (approximately $1/137$) and \(m\) as the mass of the electron or positron, yielding \(\mathcal{E}^{b}_{00} \approx -mc^2\frac{\alpha_{c}^2}{4} =-6.803\) eV. For situations where \(s \neq 0\), the energy spectrum (\ref{eq7}) shows that even the non-relativistic energy, which scales with \(\alpha_{c}^2\), reflects the effect of spacetime topology through \(\tilde{s} = s / \alpha\). The impact of spacetime topology on binding energy levels is illustrated in Figure \ref{fig:1} for \(s=1\). Moreover, we demonstrated that the energy spectra given by Eq. (\ref{eq7}) include only $s^2$ terms, indicating that the \(s=\pm 1\) quantum states are degenerate when \(\mathcal{B}_{0}=0\). Additionally, we derived exact energy spectra for Weyl pairs ($m=0$) subject to a Coulomb-type potential in the presence of an external uniform magnetic field:
\begin{equation*}
\mathcal{E}_{ns}(\alpha) = \pm 2 \frac{\hbar c}{\ell_{\mathcal{B}}} \sqrt{\alpha} \left[n + \frac{1}{2} + \tilde{s} + \frac{\sqrt{4 + \alpha^{2}_{c} + 16\tilde{s}^2}}{4}\right]^{1/2},
\end{equation*}
where \( \ell_{\mathcal{B}} = \sqrt{\frac{\hbar c}{e \mathcal{B}_{0}}} \) represents the magnetic length. This spectrum can be relevant for interacting charge carriers in magnetized monolayer materials, with \( c \) replaced by \( v_{F} \) and \(\alpha_{c}\) replaced by \(\alpha_{c} / \epsilon_{eff}\), where \( v_{F} \) is the Fermi velocity (approximately \( c/300 \)) and \(\epsilon_{eff}\) is the effective dielectric constant of the surrounding medium. For the special case where \(\alpha_{c} = 0\), \(\alpha = 1\), and \(s = 0\), the energy spectrum simplifies to:
\begin{equation*}
\mathcal{E}_{n} = \pm 2 \frac{\hbar v_{F}}{\ell_{\mathcal{B}}}  \sqrt{n + 1},
\end{equation*}
which represents the exact Landau levels for a spinless Weyl pair (non-interacting) in a flat monolayer graphene sheet. 

It is important to highlight that obtaining a complete set of energy levels in closed form to observe the Zeeman effect on positronium-like systems ($m\neq 0$) might not be feasible, regardless of whether \(\alpha=1\) or not. Exploring analytically permissible solutions for this system continues to be a promising direction for future research. This involves finding analytical solutions to the following wave equation:
\begin{equation*}
\partial^{2}_{r}\xi_{1}+\frac{2}{r}\partial_{r}\xi_{1}+\left[\frac{\varpi^2-m^{2}_{\dagger}(r)}{4}-\frac{4\left(s+\mathcal{B}r^2\right)^{2}}{\alpha^2\,r^2}\right]\xi_{1}=0,\label{WE}
\end{equation*}
where $\varpi=\omega/c$ (with $\omega$ being the relativistic frequency), $m_{\dagger}(r)=2\tilde{m}-\frac{\alpha_c}{r}$, $\tilde{m}=\frac{mc}{\hbar}$, and $\mathcal{B}=\frac{e\mathcal{B}_{0}}{8\hbar c}$.

\section*{\small{Credit authorship contribution statement}}

\textbf{Abdullah Guvendi}: Conceptualization, Methodology, Formal Analysis, Writing – Original Draft, Investigation, Visualization, Writing – Review and Editing.\\
\textbf{Omar Mustafa}: Conceptualization, Methodology, Formal Analysis,  Writing – Original Draft, 
Investigation, Writing – Review and Editing. 

\section*{\small{Data availability}}

This manuscript has no associated data.

\section*{\small{Conflicts of interest statement}}

No conflict of interest declared by the authors.

\section*{\small{Funding}}

No fund has received for this study.

\end{document}